%% file: 2020-arxiv-dp-crown.tex
\documentclass[sigconf]{acmart}

\usepackage{latexsym}
\usepackage{graphicx}
\graphicspath{{./images/}}
\usepackage{booktabs} 
\usepackage{color}  
\usepackage{amsmath}  
\usepackage{subcaption}
\usepackage{caption}
\usepackage{tikz}
\usepackage{colortbl} 
\usepackage{framed}
\usepackage{multirow}
\usepackage{multicol}
\usepackage{hyperref}
\usepackage{url}
\usepackage{balance}
\usepackage{verbatim}
\usepackage{cancel}
\usepackage{xspace} 
\usepackage[ruled,vlined]{algorithm2e}
\usepackage{bbold} 

\newcommand{\crown}{\textsc{Crown}\xspace}

\newenvironment{Snugshade}[1][236,236,236]{
    \setlength{\itemsep}{0pt}
     \setlength{\parsep}{0pt}
     \setlength{\topsep}{0pt}
     \setlength{\partopsep}{0pt}
     \setlength{\leftmargin}{1.5em}
     \setlength{\labelwidth}{0em}
     \setlength{\labelsep}{0em} 
\setlength{\parskip}{0pt}
    \definecolor{shadecolor}{RGB}{#1}%
    \begin{snugshade}%
}{%
    \end{snugshade}%
}

\newcommand{\utterance}[1]{\textit{#1}}

\setcopyright{acmcopyright}

\copyrightyear{2020}
\acmYear{2020}
\setcopyright{acmcopyright}
\acmConference[SIGIR '20]{43rd International ACM SIGIR Conference on Research and Development in Information Retrieval}
{July 25--30, 2020}{Xi'an, China}
\acmBooktitle{43rd International ACM SIGIR Conference on Research and Development in Information Retrieval
	(SIGIR '20), July 25--30, 2020, Xi'an, China}
\acmPrice{15.00}

\begin{document}
	

\title[Conversational Question Answering over Passages]
{Conversational Question Answering over Passages\\
	by Leveraging Word Proximity Networks}

\author{Magdalena Kaiser}
\affiliation{%
	\institution{Max Planck Institute for Informatics
		\\ Saarland Informatics Campus
		\\ Saarbr\"{u}cken, Germany}
}
\email{mkaiser@mpi-inf.mpg.de}

\author{Rishiraj Saha Roy}
\affiliation{%
	\institution{Max Planck Institute for Informatics
	\\ Saarland Informatics Campus
	\\ Saarbr\"{u}cken, Germany}
}
\email{rishiraj@mpi-inf.mpg.de}

\author{Gerhard Weikum}
\affiliation{%
	\institution{Max Planck Institute for Informatics
		\\ Saarland Informatics Campus
		\\ Saarbr\"{u}cken, Germany}
}
\email{weikum@mpi-inf.mpg.de}

{\tiny }\renewcommand{\shortauthors}{Kaiser et al.}

\newcommand\BibTeX{B{\sc ib}\TeX}

\newcommand{\squishlist}{
 \begin{list}{$\bullet$}
  { \setlength{\itemsep}{0pt}
     \setlength{\parsep}{3pt}
     \setlength{\topsep}{3pt}
     \setlength{\partopsep}{0pt}
     \setlength{\leftmargin}{1.5em}
     \setlength{\labelwidth}{1em}
     \setlength{\labelsep}{0.5em} } }

\newcommand{\squishend}{
  \end{list}  }

\input{sections/abstract}



\maketitle

\input{sections/introduction}
\input{sections/method}
\input{sections/architecture}
\input{sections/demo}
\input{sections/conclusion}


\bibliographystyle{ACM-Reference-Format}
\bibliography{crown}

\end{document}

%% file: sections/abstract.tex
\begin{abstract}
Question answering (QA) over text passages is a problem of long-standing interest
in information retrieval. Recently, the conversational setting has attracted
attention, where a user asks a sequence of questions to satisfy her information 
needs around a topic. While this setup is a natural one and similar to
humans conversing with each other, it introduces two key research challenges:
understanding the context left implicit by the user in follow-up questions,
and dealing with ad hoc question formulations.
In this work, we demonstrate
\crown (\textbf{C}onversational passage ranking by \textbf{R}easoning \textbf{O}ver
\textbf{W}ord \textbf{N}etworks): an unsupervised yet effective system for
conversational QA with passage responses, that supports several modes of
context propagation over multiple turns. To this end, \crown first builds
a word proximity network (WPN)
from large corpora to store statistically significant term co-occurrences.
At answering time, passages are ranked by a combination
of their similarity to the question, and coherence of query terms within:
these factors are measured by reading off node and edge weights from the WPN.
\crown provides an interface that is both intuitive for end-users,
and insightful for experts for reconfiguration to individual setups.
\crown was evaluated on TREC CAsT data, where it achieved above-median 
performance in a pool of neural methods.
\end{abstract}

%% file: sections/introduction.tex
\section{Introduction}
\label{sec:introduction}

\textbf{Motivation.} \textit{Question answering} (QA)
systems~\cite{lu2019answering}
return direct 
answers to natural 
language queries, in contrast to the standard practice of document responses.
These crisp answers are aimed at reducing users' effort in searching for
relevant information, and may be in the form of short
text passages~\cite{cohen2018wikipassageqa},
sentences~\cite{yang2015wikiqa},
phrases~\cite{quac},
or entities from a knowledge graph~\cite{huang2019knowledge}.
In this work, we deal with passages: such
\textit{passage retrieval}~\cite{salton1993approaches} 
has long been an area 
of focus for research in information retrieval (IR), and is tightly coupled
with traditional text-based QA~\cite{tellex2003quantitative}. 
Passages are one of
the most flexible answering modes, being able to satisfy both objective (factoid)
and subjective (non-factoid) information needs succinctly.

Of late, the rise of voice-based personal assistants~\cite{crook2018conversational}
like Siri, Cortana, Alexa, or the Google Assistant has drawn attention to
the scenario of
\textit{conversational question 
	answering} (ConvQA)~\cite{qu2019attentive,christmann2019look}. 
Here, the user, instead of a one-off query, fires a series of questions to the system on a topic of interest.
Effective passage retrieval often holds the key to satisfying such responses, as
short passages or paragraphs are often the most that can be spoken out loud, or 
displayed in limited screen real estate, without sacrificing coherence.
The main research challenge brought about by this shift
to the conversational paradigm is to resolve the unspoken context of the follow-up
questions. Consider our running example conversation below, that
is a mix of factoid (turns 1, 2 and 3), and non-factoid questions
(turns 4 and 5). Answers shown are excerpts from top paragraphs retrieved by
an ideal passage-based ConvQA system.
\vspace*{0.2cm}

\begin{Snugshade}
\noindent	\textbf{Question (Turn 1):} \utterance{when did nolan make his batman movies?} 
	
\noindent	\textbf{Answer:} \textit{Nolan launched one of the Dark Knight's most successful eras with Batman Begins in 2005, The Dark Knight in 2008, and the final part of the trilogy The Dark Knight Rises in 2012.}\\
	
\noindent	\textbf{Question (Turn 2):} \utterance{who played the role of alfred?}	

\noindent	\textbf{Answer:} \textit{... a returning cast: Michael Caine as Alfred Pennyworth...}\\
	
\noindent \textbf{Question (Turn 3):} \utterance{and what about harvey dent?}

\noindent 	\textbf{Answer:} \textit{The Dark Knight featured Aaron Eckhart as Harvey Dent.}\\
	
\noindent 	\textbf{Question (Turn 4):} \utterance{how was the box office reception?}

\noindent 	\textbf{Answer:} \textit{The Dark Knight earned $534.9$ million in North America and $469.7$ million in other territories for a worldwide total of $1$ billion.}\\
	
\noindent 	\textbf{Question (Turn 5):} \utterance{compared to Batman v Superman?}	
	
\noindent 	\textbf{Answer:} \textit{Outside of Christopher Nolan's two Dark Knight movies, Batman v Superman is the highest-grossing property in DC's bullpen.}	
	
\end{Snugshade}
\vspace*{0.2cm}

This canonical conversation illustrates implicit context in follow-up questions. 
In turn 2, the \textit{role of alfred} refers to the one in Nolan's Batman trilogy;
in turn 3, \textit{what about} refers to
the \textit{actor playing the role} of Harvey Dent
(the Batman movies remain as additional context all through turn $5$);
in turn 5, \textit{compared to} alludes to the \textit{box office reception}
as the point of comparison. Thus, ConvQA is far more than coreference
resolution and question completion~\cite{kumar2017incomplete}.

\textbf{Relevance.} Conversational QA lies under the general umbrella of 
\textit{conversational search},
that is of notable contemporary interest in the IR community. This is evident
through recent forums like the TREC Conversational Assistance Track
(CAsT)\footnote{\url{http://www.treccast.ai/}}~\cite{dalton19cast},
the Dagstuhl seminar on Conversational Search\footnote{\url{https://bit.ly/2Vf2iQg}}~\cite{anand2020conversational},
and the ConvERSe workshop at WSDM 2020 on Conversational Systems for 
E-Commerce\footnote{\url{https://wsdm-converse.github.io/}}.
Our proposal \crown  was originally a submission to TREC CAsT, where it 
achieved above-median performance on the track's evaluation data and
outperformed several neural methods.

\textbf{Approach and contribution.} Motivated by the lack of
a substantial volume of training data for this novel task,
and the goal of devising a lightweight and efficient system, we developed
our \textit{unsupervised} method
\crown (\textbf{C}onversational passage ranking by \textbf{R}easoning \textbf{O}ver
\textbf{W}ord \textbf{N}etworks)
that relies on the flexibility of weighted graph-based models.
\crown first builds a backbone graph referred to as
a
\textit{Word Proximity Network (WPN)}
that stores word association scores estimated from large passage
corpora like MS MARCO~\cite{nguyen2016ms}
or TREC CAR~\cite{dietz2017trec}. Passages from a baseline model
like Indri are then re-ranked according to their 
\textit{similarity} weights to question terms 
(represented as node weights in the WPN),
while preferring those passages that contain term pairs deemed significant from
the WPN, close by. Such \textit{coherence} is determined by using edge weights from
the WPN. 
\textit{Context is propagated} by various models of
(decayed) weighting of words from previous turns.

\crown enables
conversational QA over
passage corpora in a clean and intuitive UI,
with interactive response times.
As far as we know, this is the first public and open-source demo
for ConvQA over passages.
All our material is publicly available at:
\squishlist
\item \textbf{Online demo:} \url{https://crown.mpi-inf.mpg.de/}
\item \textbf{Walkthrough video:} \url{http://qa.mpi-inf.mpg.de/crownvideo.mp4}
\item \textbf{Code:} \url{https://github.com/magkai/CROWN}.
\squishend

%% file: sections/method.tex
\section{Method}
\label{sec:method}

\subsection{Building the Word Proximity Network}
\label{subsec:wpn}

Word co-occurrence networks built from large corpora have been widely 
studied~\cite{graph,dorogovtsev2001language}, and applied in many areas, like query intent
analysis~\cite{roy2011complex}.
In such networks, nodes are distinct words, and edges represent
significant co-occurrences between words in the same sentence.
In \crown, we use \textit{proximity} within a \textit{context} window,
and not simple co-occurrence: hence the term Word Proximity Network (WPN).
The intuition behind the WPN construction is to measure the coherence
of a passage w.r.t. a question, where we define coherence by words
appearing in close proximity, computed in pairs. We want to limit such
\textit{word pairs} to only those that matter, i.e., have been observed significantly
many times in large corpora. This is the information stored in the WPN.


Here, we use NPMI (normalized Pointwise
Mutual Information) for word association:
$npmi(x, y) = \log \frac{p(x,y)}{p(x) \cdot p(y)}/{-log_2 p(x,y)}$
where $p(x, y)$ is the joint probability distribution and $p(x) , p(y)$ are
the individual unigram distributions of words $x$ and $y$
(no stopwords considered). 
The NPMI value
is used as edge weight between the nodes that are similar
to conversational query tokens (Sec.~\ref{subsec:cq}). Node weights
measure the similarity between conversational query tokens and
WPN nodes appearing in the passage. 

\begin{figure}[t]
	\centering
	\includegraphics[width=0.9\columnwidth]{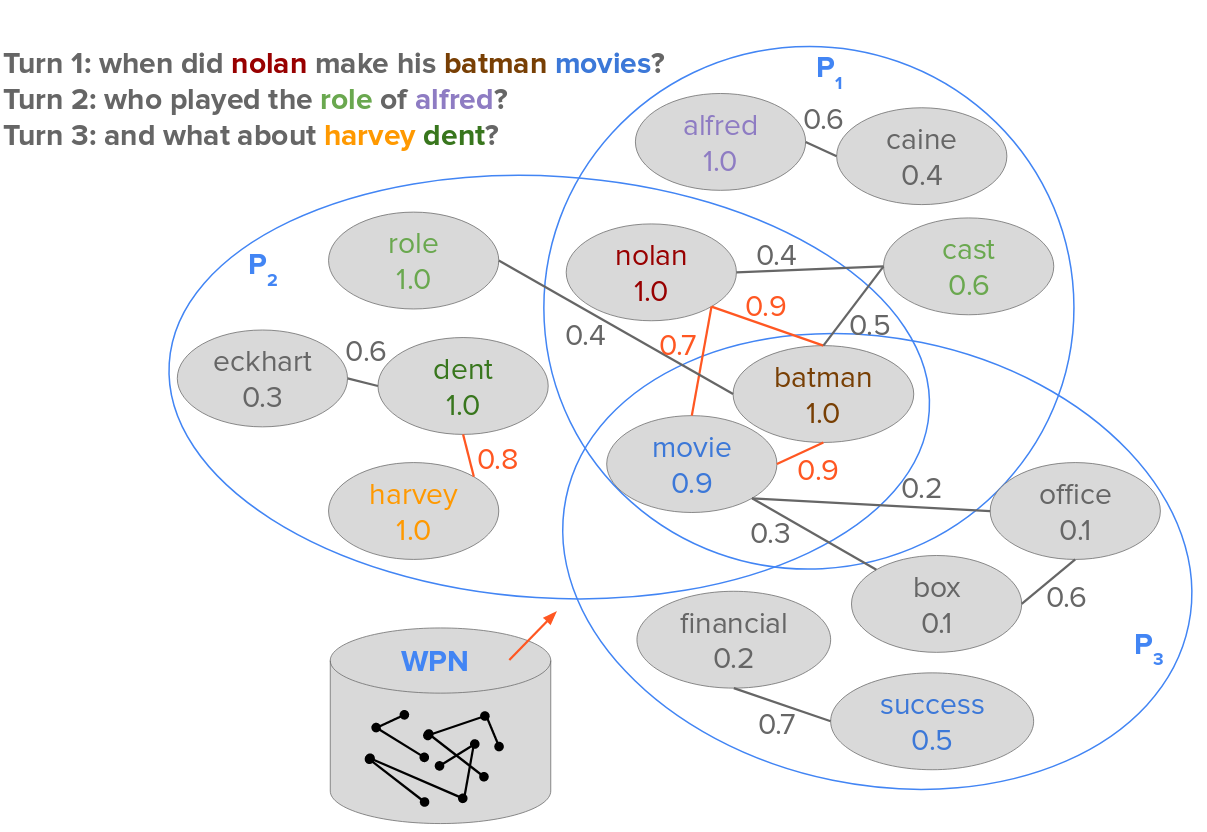}
	\caption{Sample conversation and word proximity network.}
	\label{crownNetwork}
	\vspace*{-0.5cm}
\end{figure}

Fig.~\ref{crownNetwork} shows the first three turns of
our running example, 
together with the associated fragment (possibly disconnected as irrelevant
edges are not shown) from the WPN.
Matching colors indicate which of the query words is closest to
that in the corresponding passage. 
For example, \textit{nolan} 
has a direct match in the first turn, giving it a node weight
(compared using \textit{word2vec} cosine similarity)
of $1.0$.
If this similarity is below a threshold, then the corresponding
node
will not be considered further (\textit{caine} or \textit{financial}, greyed out). 
Edge weights are shown as edge labels, considered only if they
exceed a certain threshold:
for instance, the pairs (\textit{batman},  \textit{movie}) 
and (\textit{harvey}, \textit{dent}), 
with NPMI $\geq 0.7$, qualify here. 
These edges are highlighted in orange. 
Edges like (\textit{financial}, \textit{success}) with
weight above the threshold are not considered as they are irrelevant
to the input question (low node weights), even though they appear in the given passages.






\subsection{Formulating the Conversational Query}
\label{subsec:cq}

To propagate context, \crown expands the query at a given turn $T$ using three possible
strategies to form a conversational query $cq$.
$cq$ is constructed from previous query turns $q_t$
(possibly weighted with $w_t$) seen so far:
\squishlist
	\item Strategy $cq_1$ simply concatenates the current query $q_T$ and $q_1$. No weights are used. 
	\item $cq_2$ concatenates $q_T$, $q_{T-1}$ and $q_1$, where each component has a weight
	$w_{1} = 1.0, w_{T} = 1.0, w_{T-1} = (T-1)/T$.
	\item $cq_3$ concatenates all previous turns with decaying weights ($w_t = t/T$), except for the first and
	the current turns ($w_1 = 1.0, w_T = 1.0$), as they are usually relevant to the full conversation.
\squishend
This $cq$ is first passed through Indri to retrieve a set of candidate passages,
and then used for re-ranking these candidates (Sec.~\ref{subsec:scoring}).

\subsection{Scoring Candidate Passages}
\label{subsec:scoring}

The final score of a passage $P_i$ consists of several components
that will be described in the following text. 

\noindent \textbf{Estimating similarity.}
Similarity is computed using
node weights: 
\[score_{node}(P_i) = \sum_{j=1}^n \mathbb{1}_{C_1}(p_{ij}) \cdot \max_{k \in q_t, q_t \in cq} (sim(vec(p_{ij}), vec(cq_{k})) \cdot w_t)\]
where $\mathbb{1}_{C_1}(p_{ij})$ is $1$ if the condition $C_1(p_{ij})$ is satisfied, else $0$ (see below for a definition of $C_1$). 
$vec(p_{ij})$ is the \textit{word2vec} vector of the $j^{th}$ token in the $i^{th}$ passage;
$vec(cq_{k})$ is the corresponding vector of the $k^{th}$ token in the conversational query
$cq$ and $w_t$ is the weight of the turn in which the $k^{th}$ token appeared;
$sim$ denotes the cosine similarity between the passage token and the query token embeddings.
$C_1(p_{ij})$  is defined as
$C_1(p_{ij}) := \exists cq_{k} \in cq: sim(vec(p_{ij}), vec(cq_{k})) > \alpha$
which means that condition $C_1$ is only fulfilled if the similarity between a query and a passage word
is above a threshold $\alpha$. 

\noindent \textbf{Estimating coherence.}
Coherence is calculated using edge weights:
$$score_{edge}(P_i)= \sum_{j=1}^n \sum_{k=j+1}^{W}  \mathbb{1}_{C_{21}(p_{ij}, p_{ik}), C_{22}(p_{ij}, p_{ik})} \cdot NPMI(p_{ij}, p_{ik}) $$
\begin{align*}
C_{21}(p_{ij}, p_{ik}) &:=  hasEdge(p_{ij}, p_{ik}) \land NPMI(p_{ij}, p_{ik}) > \beta \\
C_{22}(p_{ij}, p_{ik}) &:= \exists cq_{r}, cq_{s} \in cq: \\
& sim(vec(p_{ij}), vec(cq_{r})) > \alpha \\
& \land sim(vec(p_{ik}), vec(cq_{s})) > \alpha \\
& \land cq_{r} \neq cq_{s} \\
& \land \not \exists  cq_{r'}, cq_{s'} \in cq: \\
& sim(vec(p_{ij}), vec(cq_{r'})) >  sim(vec(p_{ij}), vec(cq_{r})) \\
& \lor sim(vec(p_{ik}), vec(cq_{s'})) >  sim(vec(p_{ik}), vec(cq_{s})) 
\end{align*}
$C_{21}$ ensures that there is an edge between the two tokens
in the WPN, with edge weight $> \beta$.
$C_{22}$ states that there are two words in $cq$
where one is that which is most similar to $p_{ij}$,
and the other is the most similar to $p_{ik}$.
Context window size $W$ is set to three.

\noindent \textbf{Estimating positions.} Passages with relevant sentences earlier should be preferred.
The position score of a passage is defined as:
\[score_{pos}(P_i) = max_{s_j \in P_i} \ (\frac{1}{j} \cdot (score_{node}(P_i)[s_j] + score_{edge}(P_i)[s_j]))  \]
where $s_j$ is the $j^{th}$ sentence in passage $P_i$ and $score_{node}(P_i)[s_j]$ is node score for the sentence $s_j$ in $P_i$.

\noindent \textbf{Estimating priors.} We also consider the original ranking from Indri, which can often be very useful:
$score_{indri}(P_i) = 1/rank(P_i)$ where $rank$ is the rank that the passage $P_i$ received from Indri.

\noindent \textbf{Putting it together.}
The final score for a passage $P_{i}$ consists of a weighted sum of these four individual scores:
$score(P_i) = h_{1} \cdot score_{indri}(P_i) + h_{2} \cdot score_{node}(P_i) 
+ h_{3} \cdot score_{edge}(P_i) +  h_{4} \cdot score_{pos}(P_i)$,
where $h_{1}$,  $h_{2}$, $h_{3}$ and  $h_{4}$ are hyperparameters tuned on TREC CAsT data.
The detailed method and the evaluation results of \crown are available in our 
TREC report~\cite{kaiser19crown}. General information about CAsT can be found in the TREC overview report~\cite{dalton19cast}.

%% file: sections/architecture.tex
\section{System Overview}
\label{sec:architecture}

\begin{figure}[t]
	\includegraphics[width=\columnwidth]{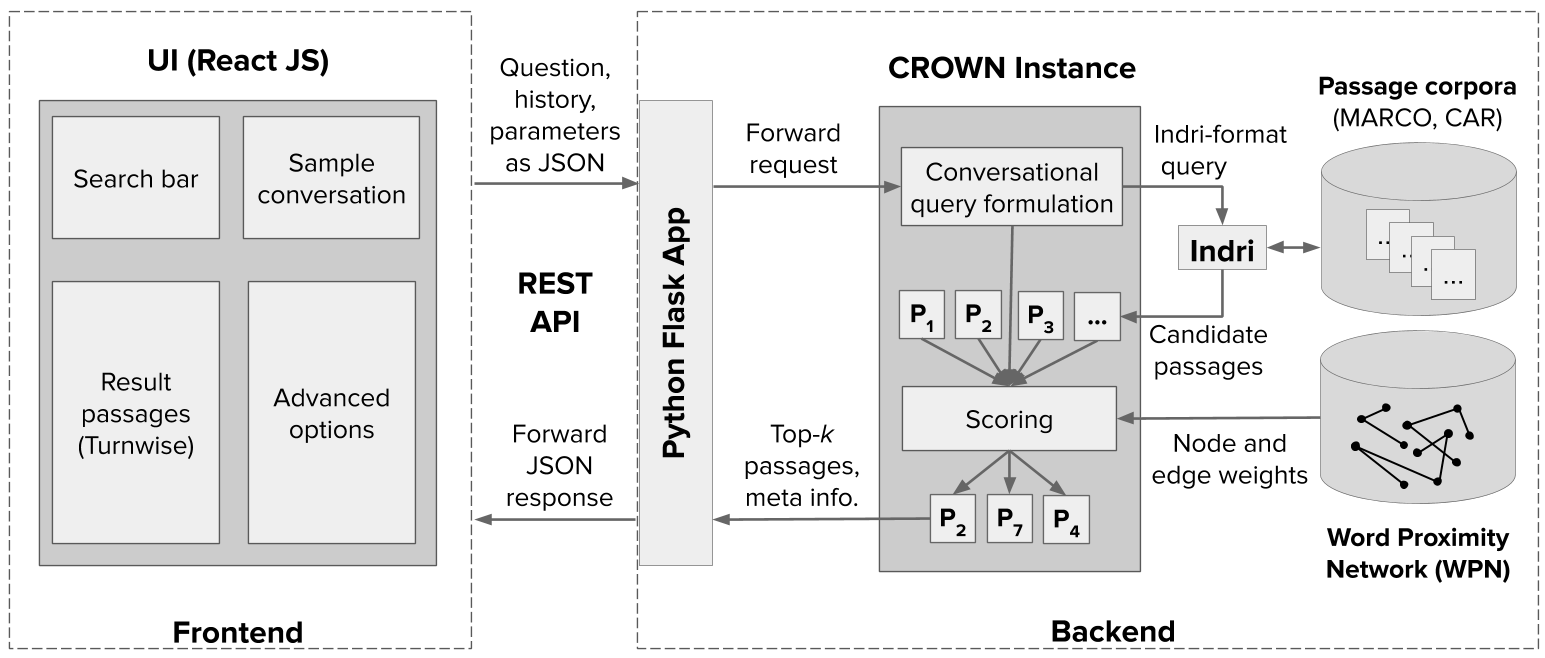}
	\caption{Overview of the \crown architecture.}
	\label{fig:architecture}
\end{figure}

An overview of our system architecture is shown in Fig.~\ref{fig:architecture}. The demo consists of a frontend and a backend, connected via a RESTful API.

\noindent \textbf{Frontend.}
The frontend has been created using the Javascript library React.
There are four main panels: the search panel, the panel containing the sample conversation, the results' panel, and the advanced options' panel. 
Once the user presses the answer button, their current question, along with the 
conversation history accumulated so far, and the set of parameters, are sent to the 
backend.
A detailed walkthrough of the UI will be presented in 
Sec.~\ref{sec:demo}.

\noindent \textbf{Backend.}
The answering request is sent via JSON to a Python Flask App, which works in a 
multi-threaded way to be able to serve multiple users. It forwards the 
request to a new CROWN instance which computes the results as described in
Sec.~\ref{sec:method}.
The Flask App sends the 
result back to the frontend via JSON, where it is displayed on the results' panel.

\noindent \textbf{Implementation Details.}
The demo requires $\simeq 170$ GB disk space and
$\simeq 20$ GB memory.
The frontend is in Javascript, and the backend is
in Python.
We used pre-trained $word2vec$ embeddings 
that were obtained via the Python library 
\textit{gensim}\footnote{\url{https://radimrehurek.com/gensim/}}. 
The Python library \textit{spaCy}\footnote{\url{https://spacy.io/}}
has been used
for tokenization and stopword removal.
As previously mentioned, 
Indri\footnote{\url{https://www.lemurproject.org/indri.php}}
has been used for candidate passage retrieval. 
For graph processing, we used the Python library 
\textit{NetworkX}\footnote{\url{https://networkx.github.io/}}. 

%% file: sections/demo.tex
\section{Demo Walkthrough}
\label{sec:demo}

\begin{figure}[!h]
	\includegraphics[width=\columnwidth]{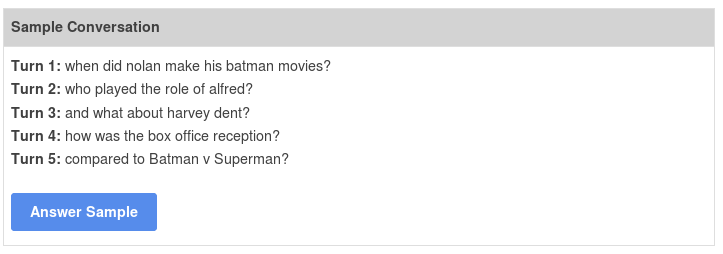}
	\caption{Conversation serving as our running example.}
	\label{fig:ui-sample}
\end{figure}

\textbf{Answering questions.} We will guide the reader through our demo using
our running example conversation
from Sec.~\ref{sec:introduction} (Fig.~\ref{fig:ui-sample}).

The demo is available at \url{https://crown.mpi-inf.mpg.de}.
One can start by typing a new question into the search bar and 
pressing \textit{Answer}, or by clicking \textit{Answer Sample} for quickly
getting the system responses for the running example.

\begin{figure}[h]
	\includegraphics[width=\columnwidth]{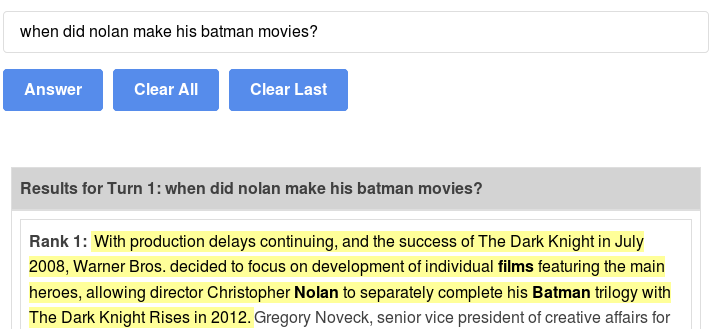}
	\caption{Search bar and rank-1 answer snippet at turn 1.}
	\label{fig:ui-question1}
\end{figure}

Fig.~\ref{fig:ui-question1} shows an excerpt from the top-ranked passage
for this first question (\utterance{when did nolan make his batman movies?}),
that clearly satisfies the information need posed in 
this turn. For quick navigation to pertinent parts of large passages, we
highlight up to three sentences from the passage (number determined
by passage length) that have the highest relevance
(again, a combination of similarity, coherence, position) to
the conversational query. In addition, important keywords are in \textbf{bold}:
these are the top-scoring nodes from the WPN at this turn.

\begin{figure}[b]
	\includegraphics[width=\columnwidth]{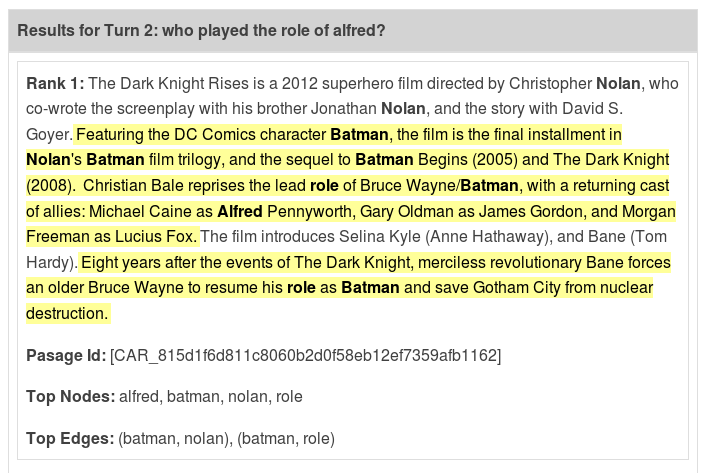}
	\caption{Top-1 answer passage for question in turn 2.}
	\label{fig:ui-question2}
\end{figure}

The search results are displayed in the answer panel below the search bar. In the default setting, the top-3 passages for a query are displayed.  
Let us go ahead and type the next question (\utterance{who played the role of alfred?})
into the input box, and explore
the results (Fig.~\ref{fig:ui-question2}).

\begin{figure}[h]
	\includegraphics[width=\columnwidth]{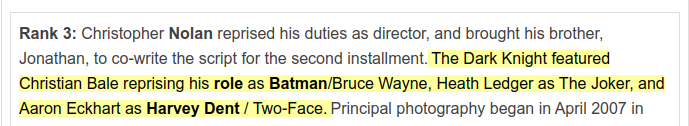}
	\caption{Answer at the 3rd-ranked passage for turn 3.}
	\label{fig:ui-question3}
\end{figure}

Again, we find that the relevant nugget of information
(\utterance{... Michael Caine as Alfred Pennyworth...}) is present in the very
first passage. We can understand the implicit context in ConvQA
from this turn, as the user does not need to specify that the role sought after
is from Nolan's batman movies. The top \textit{nodes and edges from the WPN}
are shown just after the \textit{passage id} from the corpus: 
nodes like \textit{batman} and \textit{nolan}, and edges like \textit{(batman, role)}.
These contribute to 
interpretability of the system by the end-user, and help in debugging for
the developer. We now move on to the third turn: \utterance{and what about harvey dent?},
as shown in Fig.~\ref{fig:ui-question3}.
Here, the context is even more implicit,
and the complete intent of \textit{role in nolan's batman movies} is left unspecified.
The answer is located at rank three now (see video). Similarly, we can proceed with the next two
turns. 
The result for the current question is always shown on top, while answers for previous turns do not get replaced but are shifted further down for easy reference.
In this way, a stream of (question, answer) passages is created.
Passages are displayed along with their id and the top nodes and edges found by \crown.
In the example from Figure~\ref{fig:ui-question2} not only \textit{alfred} and 
\textit{role} but also \textit{batman} and \textit{nolan}, which have been mentioned in the previous turn, are among the top nodes. 

\noindent \textbf{Clearing the buffer.} If users now want to initiate a new conversation,
they can press the \textit{Clear All} button.
This will remove all displayed answers and clear the conversation history.
In case users just want to delete their previous question (and the response),
they can use the 
\textit{Clear Last} button. This is especially helpful when exploring the effect
of the configurable parameters on responses at a given turn.

\begin{figure}[h]
	\includegraphics[width=\columnwidth]{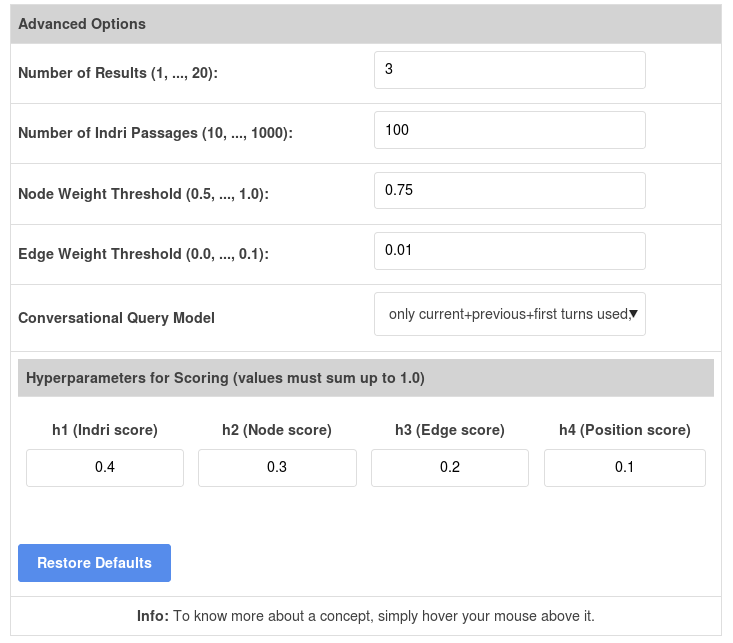}
	\caption{Advanced options for an expert user.}
	\label{fig:ui-options}
\end{figure}

\noindent \textbf{Advanced options.} An expert user can change several \crown parameters,
as illustrated in Fig.~\ref{fig:ui-options}. The first two are straightforward:
the number of top \textit{passages to display}, and to fetch from the underlying 
\textit{Indri model}. The \textit{node weight threshold} $\alpha$
(Sec.~\ref{subsec:scoring}) can be tuned depending on
the level of approximate matching desired: the higher the threshold, the more
exact matches are preferred. The \textit{edge weight threshold} $\beta$ is connected to 
the level of statistical significance of the word association measure used:
the higher the threshold, the more significant the term pair is constrained
to be. Tuning these thresholds are constrained to fixed ranges
(node weights: $0.5-1.0$; edge weights: $0.0 - 0.1$)
so as to preclude accidentally introducing a large amount of noise in the system.

\begin{figure}[t]
	\includegraphics[width=\columnwidth]{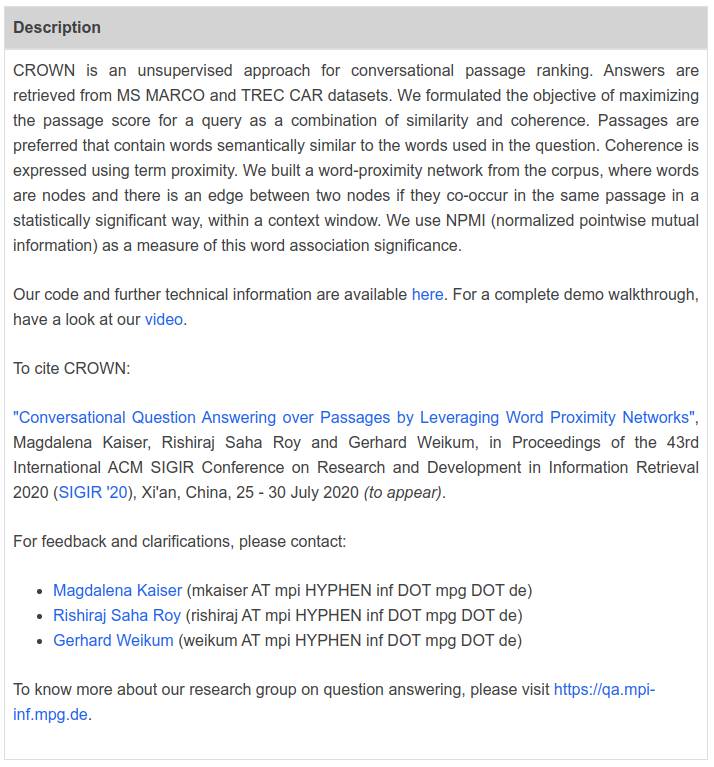}
	\caption{A summarizing description of \crown.}
	\label{fig:ui-desc}
	\vspace*{-0.6cm}
\end{figure}

The conversational query model should be selected depending upon the nature
of the conversation. If all questions are on the same topic of interest
indicated by the first question, then the intermediate turns are not so important
(select \textit{current+first turns}).
On the other hand, if the user keeps drifting from concept to concept through
the course of the conversation, then the current and the previous turn should be preferred
(select \textit{current+previous+first turns}). If the actual scenario is a mix of the two,
select \textit{all turns proportionate weights}. The first two settings may be
referred to as \textit{star and chain conversations}, respectively~\cite{christmann2019look}.
Finally, the relative weights (\textit{hyperparameters}) of the four ranking criteria 
can be configured freely ($0.0-1.0$), as long as they sum up to one. For example, 
if more importance needs to be attached to the baseline retrieval, then the Indri score
can be bumped up
(at the cost of node or edge score, say).
If
the position in the passage is something more vital, it could be raised to $0.3$, for example.
Such changes in options are \textit{reflected immediately} when a new question is asked.
Default values
have been tuned on TREC CAsT 2019 training samples.
\textit{Restore Defaults} will
reset values back to their defaults.
A brief description summarizes our contribution (Fig.~\ref{fig:ui-desc}).

%% file: sections/conclusion.tex
\section{Conclusion}
\label{sec:confut}

We demonstrated \crown, one of the first prototypes for unsupervised
conversational question answering over text passages. \crown resolves
implicit context in
follow-up questions by expanding the current query with keywords from
previous turns, and uses this new conversational query for scoring 
passages using a weighted combination of similarity,
coherence, and positions of approximate matches of query terms.
In terms of empirical performance, \crown scored above the median in 
the Conversational Assistance Track at TREC 2019, being comparable
to several neural methods.
The presented demo is lightweight and efficient, as evident in its 
 interactive response rates.
The clean UI design
makes it easily accessible for first-time users, but
contains enough configurable parameters so that experts can tune \crown
to their own setups.

A very promising extension is to incorporate
answer passages as additional context to expand follow-up questions, as users
often formulate their next questions by picking up cues from the responses
shown to them. Future work would also incorporate fine-tuned BERT 
embeddings and corpora with more information coverage.
\vspace*{-0.05cm}

%